\documentclass[12pt,letterpaper]{iopart}
\usepackage{graphicx}
\usepackage[T1]{fontenc}
\usepackage{iopams}

\begin{document}


\title{Modelling exchange bias in core/shell nanoparticles}

\author{\`{O}scar Iglesias, Xavier Batlle and Am\'{\i}lcar Labarta}
\address{Departament de F\'{\i}sica Fonamental and Institut de Nanociència i Nanotecnologia de la UB (IN$^2$UB), Universitat de Barcelona, Av. Diagonal 647, 08028 Barcelona, Spain}
\ead{oscar@ffn.ub.es, http://www.ffn.ub.es/oscar}

\begin{abstract}
We present an atomistic model of a single nanoparticle with core/shell structure that takes into account its lattice strucutre and spherical geometry, and in which the values of microscopic parameters such as anisotropy and exchange constants can be tuned in the core, shell and interfacial regions. By means of Monte Carlo simulations of the hysteresis loops based on this model, we have determined the range of microscopic parameters for which loop shifts after field cooling can be observed. The study of the magnetic order of the interfacial spins for different particle sizes and values of the interfacial exchange coupling have allowed us to correlate the appearance of loop asymmetries and vertical displacements to the existence of a fraction of uncompensated spins at the shell interface that remain pinned during field cycling, offering new insight on the microscopic origin of the experimental phenomenology.
\vspace{1pc}
\end{abstract}


\pacs{05.10 Ln,75.50.Tt,75.75.+a,75.60.-d}
\vspace{2pc}
\noindent{\it Keywords}: Monte Carlo simulation, Nanoparticles, Hysteresis, Exchange bias\\
\submitto{\JPCM}
\maketitle

\section{Introduction}

Although the observation of shifted hysteresis loops after field cooling (FC) along the field direction was first reported in nanoparticles more than five decades ago \cite{Meiklejohn_PR}, most of the research exploiting this effect have been conducted in thin film systems formed by an antiferromagnetic (AFM) layer in contact with a ferromagnet (FM) \cite{Reviews}. However, recently renewed interest in the study of exchange bias (EB) in core/shell nanoparticles has been launched by their potential use as recording media with improved thermal stability \cite{Skumryev}. A number of experiments performed in a variety of nanoparticle systems with a compound structure having a FM core surrounded by an AFM shell, formed at the outer surface of the core by oxidation of the FM structure, display common macroscopic effects that usually accompany the observation of loop shifts and that are associated to the appearance of the EB effect \cite{Nogues_physrep05}. Among these, the microscopic origin of the coercivity increase, particle size and shell thickness dependence, training effects, dependence on the cooling field magnitude and sign, hysteresis loop asymmetries and vertical shifts still deserve a proper explanation from the theoretical point of view \cite{Iglesias_JNN07}. Despite the ressemblance between the core/shell structure of the nanoparticles and the FM/AFM composition of thin film bilayers, the existent theoretical models for the former systems \cite{Kiwi_jmmm01,Stamps_jpd00} may not be not well suited for particle systems. 
In the case of layered systems, ideal interfaces can be considered to be fully compensated or non-compensated depending on the nature of the interfacial coupling and roughness caused by defects and imperfections are usually invoked to account for the magnitude of the measured EB fields. 
In contrast, the nature of the core/shell interface in nanoparticles is given by surface effects associated to the finite size of the particle and by the particular geometry of the lattice. All these peculiarities made necessary the use of simulation methods like Monte Carlo that are able to account for the magnetic order at the interfaces with atomistic detail, so as to clarify the microscopic origin of the phenomenology associated to EB effects \cite{Iglesias_JNN07,Iglesias_prb05,Iglesias_phab06,Iglesias_jmmm07}.

\section{Model and simulation method}

In order to understand the origin of the EB phenomenology, we have performed Monte Carlo simulations based on a classical spin model of a core/shell nanoparticle with different properties at the core, shell and interfacial regions. The particles are spheres of total radius $R$ with a core of radius $R_C$ and constant shell thickness $R_{Sh}=3$a ($a$ is the lattice constant) enclosing $N$ spins placed at the nodes of a simple cubic lattice. The corresponding Hamiltonian for the atomic spins ${\vec S}_i$ is given by:  
\begin{eqnarray}
\label{Eq1}
{ H}/k_{B}= 
-\sum_{\langle  i,j\rangle}J_{ij}{\vec S}_i \cdot {\vec S}_j   
-k_C\sum_{i\in \mathrm{C}}(S_i^z)^2
-k_S\sum_{i\in \mathrm{Sh}}(S_i^z)^2  	
-\sum_{i= 1}^{N} \vec h\cdot{\vec S_i}
\end{eqnarray}
where $\vec{h}$ is the magnetic field applied along the easy-axis direction with module $\vec{h}=\mu\vec{H}/k_B$ in temperature units ($\mu$ is the magnetic moment of the spin). 
In the first term the exchange constants at the different regions are FM at the core $J_{\mathrm{C}}= 10$ K, AF at the shell ($J_{\mathrm{S}}= -5$ K) and for the interfacial spins [those in the core (shell) having neighbours in the shell (core)], we will consider AF coupling $J_{\mathrm{Int}}< 0$ of varying strength. The second and third terms correspond to the uniaxial anisotropy energy, with $k_C= 1$ and $k_S$ the anisotropy constants for core and shell spins respectively. The last value will be varied from 1 to 10. The simulations have been performed using the Monte Carlo method with the standard Metropolis algorithm for continuous spins with single spin flip dynamics (see \cite{Iglesias_prb05} for details). The hysteresis loops have been computed cycling the magnetic field between $h= \pm 4$ K, after cooling from a temperature higher than the FM core ordering temperature down to $T=0.1$ in the presence of a magnetic field $h_{FC}= 4$ K. 
The detailed shape of the loops depend of course on the rate of change of the magnetic field along the loop. Our simulations have been performed by changing the magnetic field in steps $\delta h= 0.1$ K and averaging the magnetization components during 100 MC steps, discarding the initial 100 MCS after every field step. Due to our specific choice of trial step in the MC dynamics, the shape of the hysteresis loops does not show any appreciable sweeping rate dependence, only the values of the coercive fields vary but the conclusions regarding the EB effects remain unaltered. 

\begin{figure}[t] 
\centering 
\includegraphics[width=0.5\textwidth]{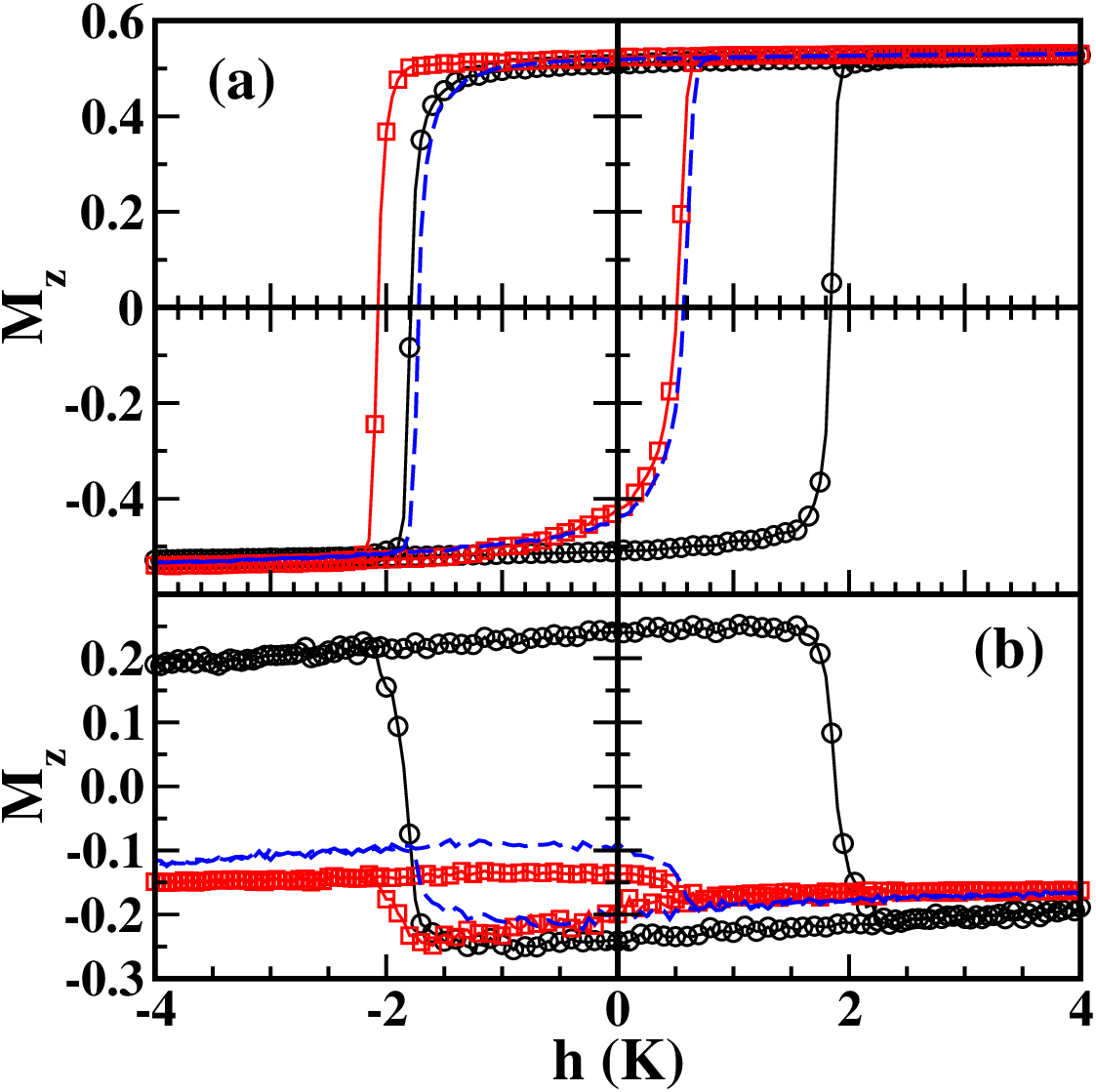}
\caption{(Colour online) Hysteresis loops after cooling in a magnetic field $h_{FC}= 4$ for a particle with $R_{Sh}= 3$a and radius $R= 12$a for different values of the shell anisotropy constant $k_{\mathrm S}= 1$ (black circles), $k_{\mathrm S}= 5$ (dashed blue line),  $k_{\mathrm S}= 10$ (red squares) and $k_C= 1$. The values of the exchange couplings are: $J_{\mathrm C}= 10$, $J_{\mathrm S}= -0.5 J_{\mathrm C}$, and $J_{\mathrm {Int}}= -0.5$. Panel (a) shows the total magnetization along the field direction, while panel (b) shows only the contribution corresponding to the interfacial spins at the shell.
}
\label{Fig1_fig}
\end{figure}

\section{Results}
\begin{figure}[t] 
\centering 
\includegraphics[width=0.5\textwidth]{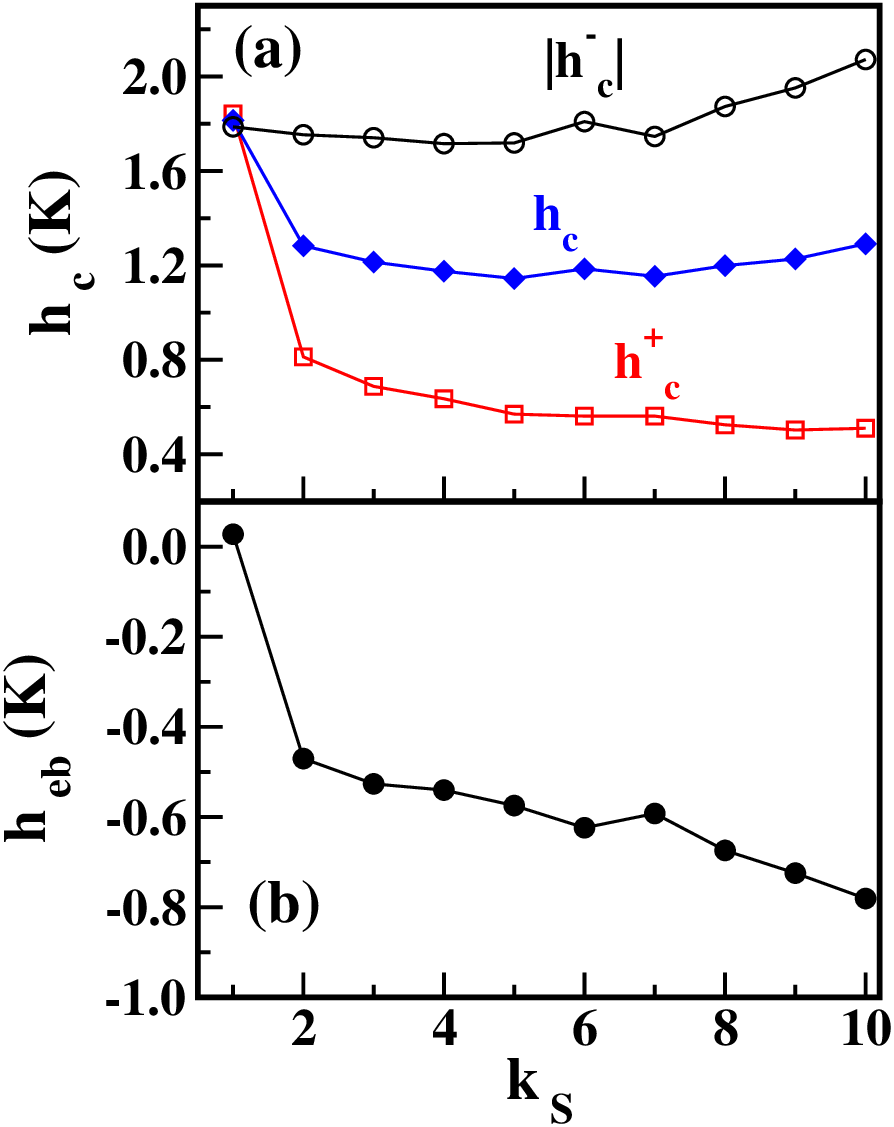}
\caption{(Colour online) Dependence of the coercive fields at the descreasing and increasing field branches $h^-_c$, $h^+_c$, coercive field $h_c$ (a panel) and exchange bias field $h_{eb}$ (b panel) on the value of the of the surface anisotropy $k_S$ for a spherical particle with $R= 12$, $R_{Sh}= 3$ and $k_C= 1$.
}
\label{Fig2_fig}
\end{figure}

We have first investigated the influence of the strength of the shell anisotropy constant $k_S$ on the observation of shifted loops for a given particle size $R= 12 a$.  
As it is also the case for bilayered thin films, a necessary condition to obtain shifted hysteresis loops after field cooling from above the Ne\'el temperature is that there must be a part of the spins in the AFM shell that are pinned along the cooling field direction during the field cycling \cite{Reviews}. Since the FM core is supposed to reverse during the hysteresis loop, one may argue that the shell spins at the core/shell interface may be dragged by the core spins during their reversal and that this may induce the reversal of the rest of shell spins, hindering the EB effect. 
Hysteresis loops simulated for values of $k_S$ in the range $1-10$ are displayed in Fig. \ref{Fig1_fig} for three representative values $k_S= 1, 5, 10$ where we can see that the loop for the particle with lower anisotropy is completely symmetric, while the ones for higher anisotropies (Fig. \ref{Fig1_fig}a) are shifted along the field axis towards the field cooling direction, confirming the reasoning above. 
%
From the values of the coercive fields at the decreasing and increasing field branches ($h^-_c$ and $h^+_c$), the magnitude of the coercive field $h_C$ and the loop shift $h_{eb}$ can be evaluated as $h_c= h^+_c- h^-_c$ and $h_{eb}= h^+_c+ h^-_c$. The dependence of these quantities on $k_S$ as extracted from the simulated hysteresis loops is displayed in Fig. \ref{Fig2_fig}, that shows that there is a minimal value of $k_S$ for the observance of EB. When increasing $k_S$ above this value, the bias field increases progressively as the proportion of interfacial spins pinned during the hysteresis loop increases and finally saturates. In contrast, in the presence of EB, $h_C$ is reduced with respect to the low anisotropy case, but its value does not show appreciable variations with $k_S$.  

Independently on the value of $k_S$, a net magnetization is induced at the core/shell interface after the cooling process, as indicated by the similar values attained by the net magnetization of the shell interfacial spins after the field cooling under $h_{FC}= 4$ K, as can be seen in Fig. \ref{Fig1_fig}b. These values are negative because AF coupling at the interface has been considered. However, the same figure reveals that the behavior of the shell interfacial spins is quite different in the high and low anisotropy cases. For $k_S= 1$, the loop of the interfacial shell spins is completely symmetric and their net magnetization reverses progressively when decreasing $h$. As $h$ is cycled between its maximum positive and minimum values, the net magnetization sign changes accordingly, without changing value. This indicates that the interfacial surface spins are not pinned during the hysteresis cycle and, therefore, no EB is observed. 

\begin{figure}[tbp] 
\centering 
\includegraphics[width=0.5\textwidth,angle= 0]{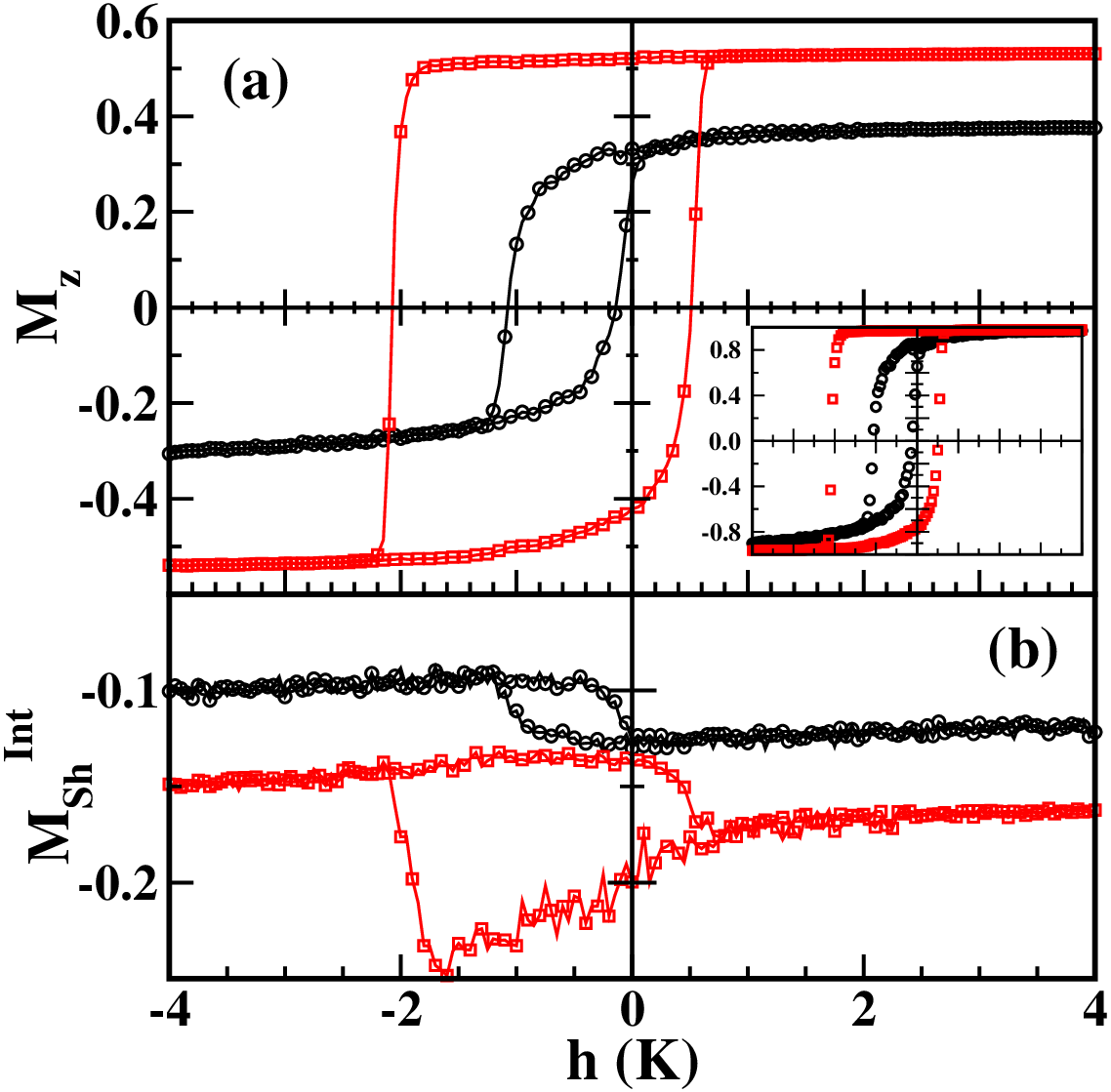}
\caption{(Colour online) Hysteresis loops after cooling in a magnetic field $h_{FC}= 4$ for two particles with $R_{Sh}= 3$a, $k_S= 10$ and different radius $R= 8$a (black circles) and $R= 12$a (red squares). Panel (a) shows the total magnetization along the field direction and the inset shows the contribution of core spins to the hysteresis loops. In panel (b), the contributions of the interfacial spins at the shell for the two particle sizes is displayed.
}
\label{Fig3_Fig}
\end{figure}

This is in contrast with the $k_S= 5, 10$ cases, for which the loop becomes asymmetric and shifted. During the descreasing branch, there is a progressive decrease of the interfacial magnetization $M^{Int}_{Sh}$ as $h$ approaches $h^-_C$, while the core magnetization remains essentially constant and reverses coherently as indicated by the sharp jump at $h^-_C$ in the Inset of Fig. \ref{Fig3_Fig}a. At the end of this branch, the magnetization is still negative, with a value slightly higher than at the beginning, which is an indication of the existence of a small fraction of pinned spins that is directly related to the value of the exchange bias field, as we showed previously \cite{Iglesias_prb05}. 
Moreover, $M^{Int}_{Sh}$ remains almost constant along the increasing branch, presenting a small decrease near $h^+_C$ due to the formation of non-uniform reversed domains in the core that propagate into the shell \cite{Iglesias_phab06,Iglesias_jmmm07} (notice the rounded shoulder in the loop of figure \ref{Fig3_Fig}a).  

Finally, we have investigated the role played by the particle size on the observation of EB by comparing simulation results for two particles with the same shell thickness $R_{Sh}= 3a$ and shell anisotropy $k_S= 10$, but total radii $R= 8a, 12a$. The only difference between the two particles is the core radius ($R_C= 5a, 9a$, respectively), so that the number of uncompensated spins at the interface will be different and, therefore, the net magnetization at the interface induced by field cooling is expected to change accordingly. 
In figure \ref{Fig3_Fig}a, we show the simulated hysteresis loops together with the contribution of the core spins (inset) and, in figure \ref{Fig3_Fig}b, the contribution of shell interfacial spins to the hysteresis loop. For both particle sizes, EB is clearly observed, although their hysteresis loops present evident differences.
First of all, for the smallest particle, both coercive fields are negative and the computed values of $h_C= 0.466$ K and $h_{eb}= -0.608$ K are significantly lower than for the biggest particle ($h_C= 1.292$ and $h_{eb}= -0.781$ K). One would expect an increase of these to quantities as the particle size decreases, due to the increasing proportion of interfacial spins \cite{Dobrynin_apl05}. However, it should be noticed that the magnitude of $h_{eb}$ depends essentially on the uncompensated magnetization \cite{Iglesias_prb05} at the core/shell interface and, as it can be checked in figure \ref{Fig3_Fig}b, after field cooling, for the particular radius $R_C=8$a, this is smaller than for the particle with $R_C= 12$a. 
This is in close relation to the observation that, also at difference from the biggest particle, for the smallest particle, both branches of the loop are rounded. This fact indicates that, for that particle, the reversal mechanism is non-uniform in both loop branches as can be further corroborated by the shape of the loop for the core contribution shown in the Inset of figure \ref{Fig3_Fig}a (circles). 
A final observation on the size effects is that the loops for both particle sizes appear shifted upward also in the direction of the magnetization axis, in agreement with several experimental results \cite{Zheng_prb04,DelBianco_prb04,Tracy_prb05,Passamani_jmmm06,Mumtaz_jmmm07}. In previous works \cite{Iglesias_phab06,Iglesias_jmmm07}, we have related the vertical loop shifts to the existence of uncompensated pinned moments at the interface, which are also responsible for the different reversal machanisms in the loop branches. We also showed that the magnitude of the vertical shift increases with the interfacial exchange coupling. Therefore, a bigger vertical shift (the difference between the saturation magnetization values at both loops branches) in the smaller particle is in agreement with the more pronounced non-uniformity in the magnetization reversal modes and with the biggest difference in the saturation values of $M^{Int}_{Sh}$.

\ack
We acknowledge CESCA and CEPBA under coordination of C$^4$ for computer facilities. This work has been supported by the Spanish MEyC through the MAT2006-03999, NAN2004-08805-CO4-01/02 and Consolider-Ingenio 2010 CSD2006-00012 projects, and the Generalitat de Catalunya through the 2005SGR00969 DURSI project. 


\end{document}